\documentclass[a4paper]{article}
\usepackage{epsfig}
\begin{document}
\newcommand{\be}{\begin{equation}}
\newcommand{\ee}{\end{equation}}
\newcommand{\bea}{\begin{eqnarray}}
\newcommand{\eea}{\end{eqnarray}}
\newcommand{\nn}{\nonumber \\}
\newcommand{\nicht}[1]{}
\newcommand{\de}{{\rm d}}
\newcommand{\ie}{{\rm i}}
\newcommand{\hr}{\hat r}
\newcommand{\hs}{\hat s}
\newcommand{\hp}{\hat\phi}
\newcommand{\hc}{\hat\chi}
\newcommand{\hC}{\hat C}
\newcommand{\hI}{\hat I}
\newcommand{\hd}{\hat{\delta}_0}
\newcommand{\ex}[1]{{\rm e}^{#1}}
\newcommand{\cs}[1]{c_{{\rm s}#1}}
\newcommand{\crr}[1]{c_{{\rm r}#1}}
\newcommand{\cp}[1]{c_{\phi #1}}
\newcommand{\cc}[1]{c_{\chi #1}}
\newcommand{\cd}{c_{\delta_0}}
\newcommand{\sd}{s_{\delta_0}}
\newcommand{\cdz}{c_{2\delta_0}}
\newfont{\Kapfont}{cmbx10 scaled 1728}

\vspace*{1cm}
\begin{center}
{\Kapfont Floating Bodies of Equilibrium I}
\end{center}
\vspace{1cm}

\begin{center}
\bf Franz Wegner, Institut f\"ur Theoretische Physik \\
Ruprecht-Karls-Universit\"at Heidelberg \\
Philosophenweg 19, D-69120 Heidelberg \\
Email: wegner@tphys.uni-heidelberg.de
\end{center}
\vspace{1cm}

\paragraph*{Abstract} A long cylindrical body of circular cross-section
and homogeneous density may float in all orientations around the cylinder
axis. It is shown that there are also bodies of non-circular cross-sections
which may float in any direction. Apart from those found by Auerbach for
$\rho=1/2$ there are one-parameter families of cross-sections for
$\rho\not=1/2$ which have a $p$-fold rotation axis. For given $p$ they exist
for $p-2$ densities $\rho$. There are strong indications, that for all $p-2$
densities one has the same family of cross-sections.

\section{Introduction}

A long standing problem asked by Stanislaw Ulam in the Scottish Book
\cite{Scottish} (problem 19) is, whether a sphere is the only solid of uniform
density which will float in water in any position. A simpler, two-dimensional,
question is: Consider a long log of circular cross-section; it will,
obviously, float in any position without tending to rotate. (Of course, the
axis of the log is assumed to be parallel to the water surface.) Are there any
additional cross-section shapes such that the log will float in any position?
This question was answered in 1938 by Auerbach \cite{Auerbach} for the special
density $\rho=1/2$ (the density of water is normalized to unity). He
showed, that for this special density there is a large variety of
cross-sections.

Here we consider the case $\rho\not= 1/2$. We observe that for special
densities one may deform the circular cross-section into one with $p$-fold
symmetry axis and mirror symmetry. There are $p-2$ densities, for which this
is possible. For given $p$ the densities appear pairwise as $\rho$ and
$1-\rho$. For odd $p$ there is also a solution for $\rho=1/2$. These solutions
can be expanded in powers of the deformation $\epsilon$. In polar coordinates
$(r,\psi)$ they are written 
\be
r(\psi) = \bar r (1+2\epsilon \cos(p\psi)+ 2\sum_{n=2}^{\infty} c_n
\cos(pn\psi)),
\ee
where the coefficients $c_n$ are functions of $\epsilon$, with
$c_n=O(\epsilon^n)$.
The associated densities depend on $\epsilon$.

I have carried through the calculations of the
coefficients up to order $\epsilon^7$ and surprisingly I obtain the same
expansion for all $p-2$ solutions, although I did expect agreement only for
pairs $\rho$ and $1-\rho$. Thus I conjecture that there is a one-parameter
family (parametrized by $\epsilon$) of bodies for given $p$ which floats
indifferently at $p-2$ different densities.

The outline of the paper is as follows: In the next section I derive some
general properties of these cross-sections. In the following section I
consider the special case $\rho=1/2$. The case of $\rho\not=1/2$ is treated in
first order of the distortion in section 4 and in higher orders in section 5.
Section 6 contains a short conclusion.

\section{General Considerations}

I start with some considerations for general densities $\rho$. I denote the
cross-section area of the log by $A$, the cross-section above the water be
$A_1$, that below the water $A_2$, then
\be
A_1 = (1-\rho) A, \quad A_2 = \rho A
\ee
according to Archimedes. Moreover I denote
by $m$ the total mass of the log,
by $m_{1,2}$ the masses above/below the water-line,
by $C_{1,2}$ the center of mass above/below the water-line, 
by $h_{1,2}$ the distance of $C_{1,2}$ from the water-line,  
and by $\hat L$ the length of the log.

Then the potential energy of the system is
\bea
V &=& m_1 g h_1 + (m-m_2) g h_2 = \rho A_1 \hat L h_1 g 
+ (A_2 \hat L-A_2 \rho \hat L) g h_2 \nn
&=& \rho (1-\rho) A \hat L g (h_1 +h_2).
\eea
Thus the difference in height between the two centers of mass, $h_1+h_2$,
has to be independent of the orientation of the log. 
This does not imply, that $h_1$ and $h_2$ are separately constant.
Moreover the line $C_1 C_2$ connecting the two centers of mass has to be
perpendicular to the water-level. In the following I will consider
convex simply connected cross-sections. This condition may be relaxed, but it
is necessary, that the water-level crosses the circumference of the log at
exactly two points denoted by $L$ and $R$. The distance between these two
points be $2l$.

\begin{center}
\epsfig{file=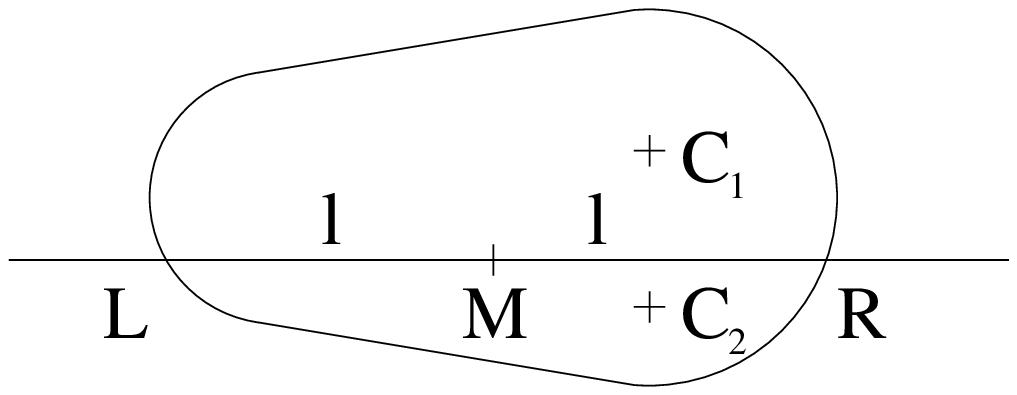,scale=0.5,angle=0}\\
Figure 1
\end{center}

\medskip

Let us consider now what happens if we rotate the log by an infinitesimal
angle $\delta \phi$. Since the area above and below the water has to be conserved the rotation is around the midpoint $M$ of the line $LR$.
Then on the right the area $l^2\delta\phi/2$ rises above the water
and on the left the same area disappears below the water-level. Let us consider
the shift of the centers of mass in horizontal direction. There are two
changes: Due to the rotation $C_1$ moves to the left by $\delta \phi h_1$
and $C_2$ to the right by $\delta \phi h_2$. The appearance of the area
$l^2 \delta \phi/2$ on the right and the disappearance of the same area 
on the left moves the center $C_1$ of mass by 
$\delta \phi (\int_0^l l^2 dl-\int_{-l}^0 l^2 dl) /A_1
= \frac 23 \delta \phi l^3 / A_1$ to the right.
Similarly the center of mass below the water level $C_2$ moves to the left by
$\frac 23 \delta \phi l^3 / A_2$.

In total the shift to the right of $C_1$ is
$(-h_1+ \frac 23 l^3 /A_1) \delta \phi$ and that of $C_2$
$(h_2 -\frac 23 l^3/A_2) \delta\phi$. Both have to be equal
since otherwise a torque would be created. Thus we have
\be
\frac 23 l^3 (\frac 1{A_1} + \frac 1{A_2}) = h_1+h_2. \label{cond}
\ee
Since the distance $h_1+h_2$ has to be constant and also $A_1$ and $A_2$
are constant, we obtain the important result,
that also $l$ is constant. If (\ref{cond}) is not fulfilled, but the 
left hand side is larger then the right hand side, then the log is in
stable equilibrium, if the left hand side is less then the right hand side
then the equilibrium is unstable. But here we are interested in an
indifferent equilibrium, where equality (\ref{cond}) holds.

When we continue to rotate the log, then in general the mid-point $M$ between 
$L$ and $R$ will move. It can move only along the direction of the line 
$LR$. We consider now the loci of the points $M$ as a function of $\phi$.
In order to do this we keep however the orientation of the log fixed and rotate
the direction of the gravitational force and the orientation of the
watersurface. Since $M$ can move only in direction of the line $LR$ whose
slope is now given by $\tan \phi$ we can parametrize the loci of $M$ by
$x_M(\phi)$ and $y_M(\phi)$, where $s(\phi) d\phi$ is the
shift of $M$ along $LR$ as we rotate by $d\phi$. It is now obvious that
the solution of the problem can be parametrized by 
\bea
x_R(\phi) = x_M(\phi) + l \cos \phi, &&
y_R(\phi) = y_M(\phi) + l \sin \phi, \label{xy} \\
x_M(\phi) = x_M(0)+\int_0^{\phi} s(\phi) \cos \phi \, \de\phi, &&
y_M(\phi) = y_M(0)+\int_0^{\phi} s(\phi) \sin \phi \, \de\phi.
\label{M}
\eea
$x_M(\phi)$ and $y_M(\phi)$ parametrize the envelope of the water-lines $LR$.
Considering the boundary of the log we observe, that the part of
the circumference which disappears on one side under rotation by $\delta\phi$
below the water-line and appears on the other side is
$\sqrt{s^2+l^2}\delta\phi$ on both sides, since in horizontal direction the
contribution is $s\de\phi$ and in vertical direction $l\de\phi$. Thus the
part of the circumference below the waterline is constant independent of the
orientation.
Obviously the envelope of the water-line must be closed, that is
\be
\int_0^{2\pi} s(\phi) \cos \phi \, \de\phi =0, \quad
\int_0^{2\pi} s(\phi) \sin \phi \, \de\phi =0 \label{int2}
\ee
It must be noted however, that the boundary of the log is
not only given by eq. (\ref{xy}), but it is also obtained by reversing 
the sign of $l$, that is not only by considering the points $R$ but also the
points $L$
\be
x_L(\chi) = x_M(\chi) - l \cos \chi, \quad
y_L(\chi) = y_M(\chi) - l \sin \chi.
\ee
The problem is now to find envelopes, parametrized by $s(\phi)$, so that 
both descriptions coincide. Of course the arguments $\phi$ and $\chi$ are
different in both expressions in order to describe the same point on the
boundary.

\section{The case $\rho=\frac 12$}

In the special case $\rho=\frac 12$ one has $A_1=A_2=A/2$, with the implication
that for $\phi$ and $\phi+\pi$ one has the same line $LR$ separating the
part of the log above and that below the water-level, only $L$ and $R$ are
exchanged, but $M$ is the same for $\phi$ and $\chi=\phi+\pi$
\be
x_M(\phi)=x_M(\phi+\pi), \quad y_M(\phi)=y_M(\phi+\pi). \label{m}
\ee
If we differentiate these equations with respect to $\phi$ using the
representation (\ref{M}) we obtain the condition
\be
s(\phi+\pi)=-s(\phi) \label{s}
\ee
and taking it for $\phi=0$ yields
\be
\int_0^{\pi} s(\phi) \cos \phi \, d\phi =0, \quad
\int_0^{\pi} s(\phi) \sin \phi \, d\phi =0 \label{int}
\ee
The conditions (\ref{s}) and (\ref{int}) and a sufficient 'convexity' is
sufficient to obtain a body which flows in each direction. Curves of this type
were already discussed by Zindler in section 6 of \cite{Zindler} but apparently
without reference to this physical problem.

As an example we consider
\be
s(\phi) = a \cos(3 \phi) \label{ex1}
\ee
which yields
\be
x_M(\phi)=\frac a4 \sin(2\phi) + \frac a8 \sin(4\phi), \quad
y_M(\phi)=\frac a4 \cos(2\phi) - \frac a8 \cos(4\phi) - \frac a8, \label{ex2}
\ee
a curve which has kinks at ($\pm 3\sqrt 3 a/16$, $a/16$), (0, $-a/2$). For 
$l>3a$ the circumference is convex, but it might be that it is sufficient
to have $l>a/2$, so that the circumference lies completely outside the
loci of $M$.

\epsfig{file=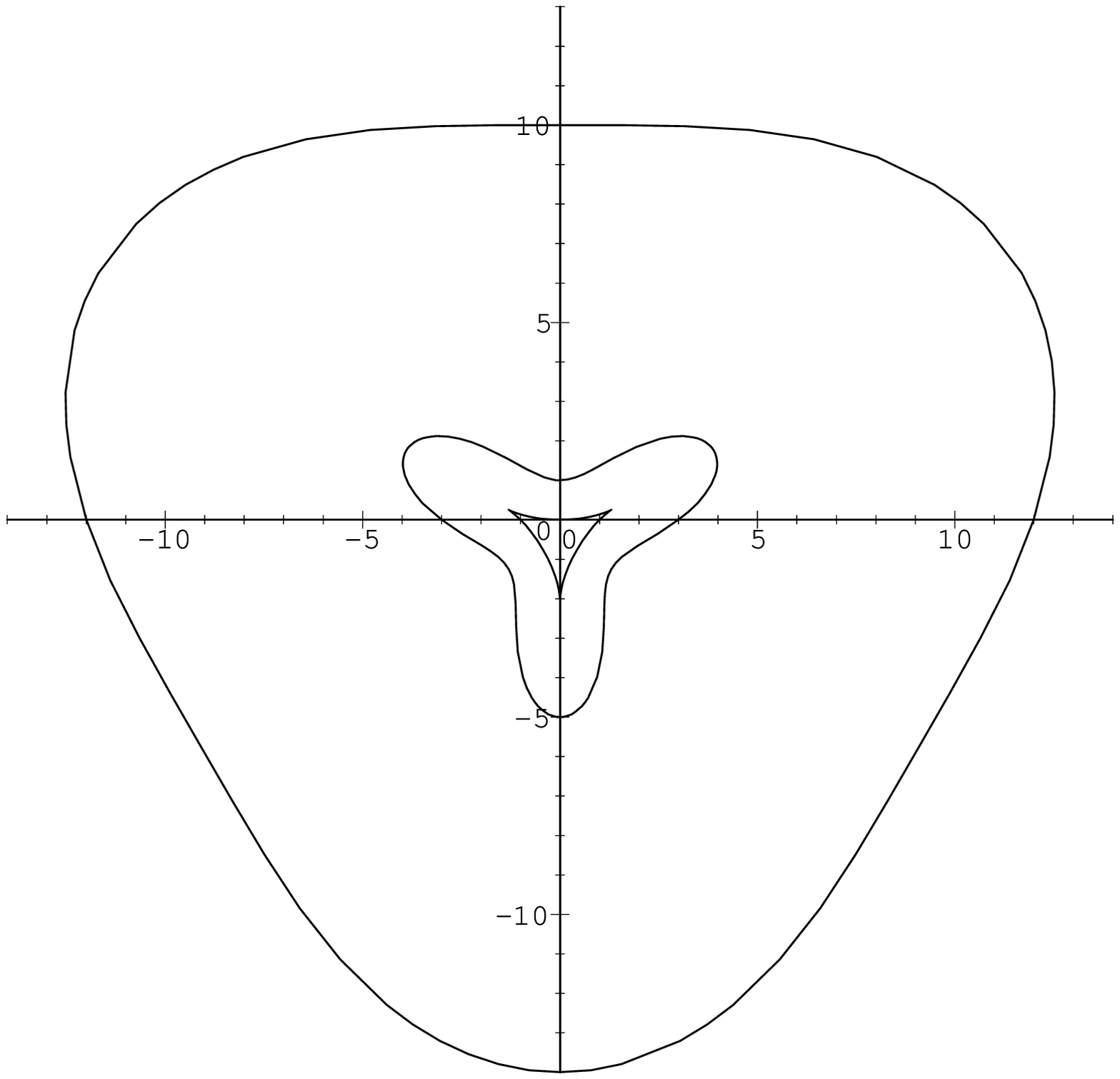,scale=0.4,angle=270}
\hfill
\parbox[t]{4.7cm}
{\vspace{1cm}
Figure 2. This figure shows the loci of M for $a=4$ (innermost curve)
and the boundaries for $l=3$ and $l=12$ (outermost curve). A log
with the cross-section of the outermost curve and $\rho=1/2$ will flow
in water in any position without tendency to rotate. The water-line will
be tangent to the innermost curve.}

One may however, choose a cross-section with very elementary pieces of the
boundary. We may e.g. choose a triangle as envelope of the waterlines. Then
we obtain a cross-section bounded by six straight lines and six arcs as shown
in the next figure. This corresponds to an $s(\phi)$ consisting of a sum of
three $\delta$-distributions.

\epsfig{file=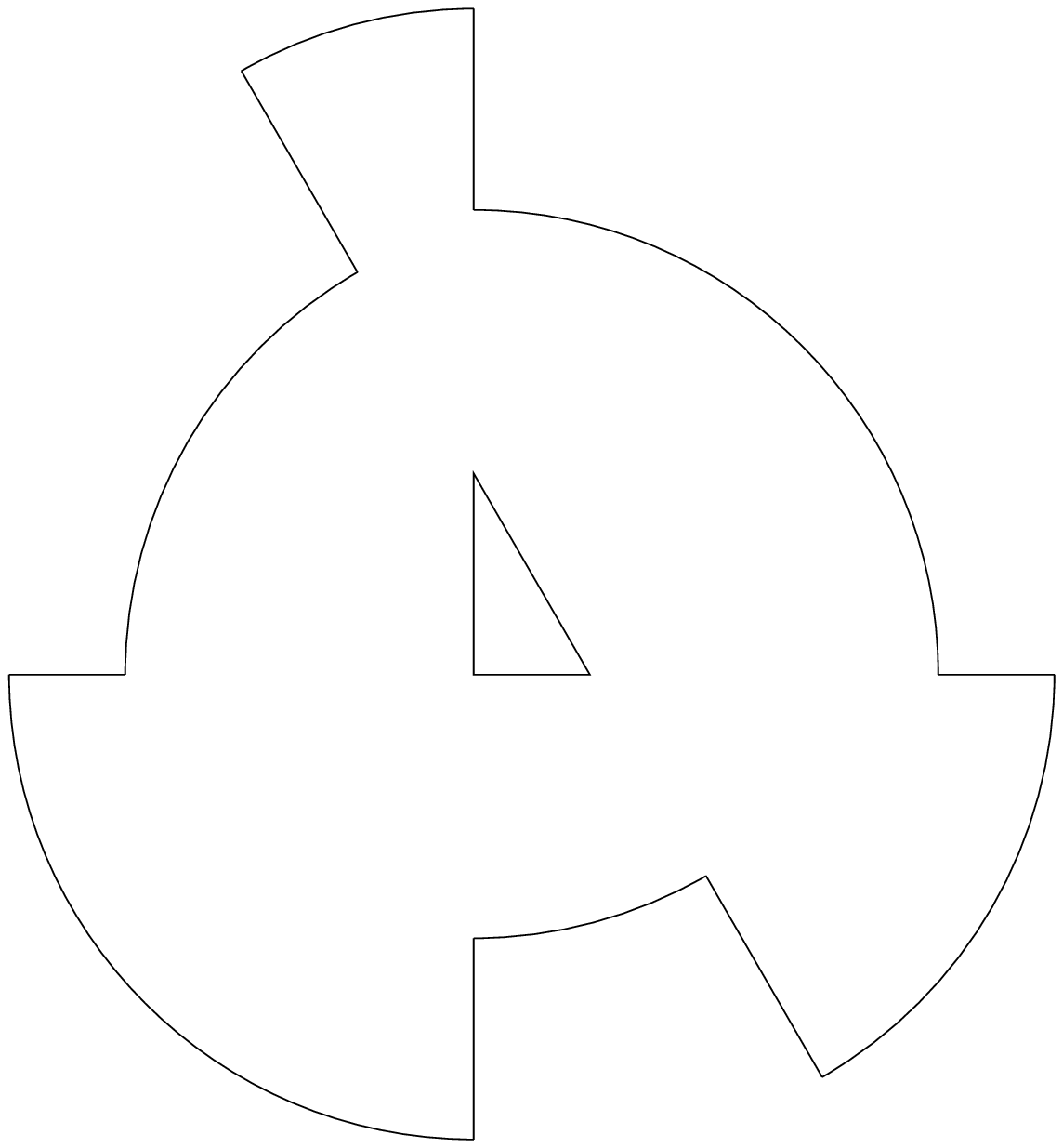,scale=0.4}
\hfill
\epsfig{file=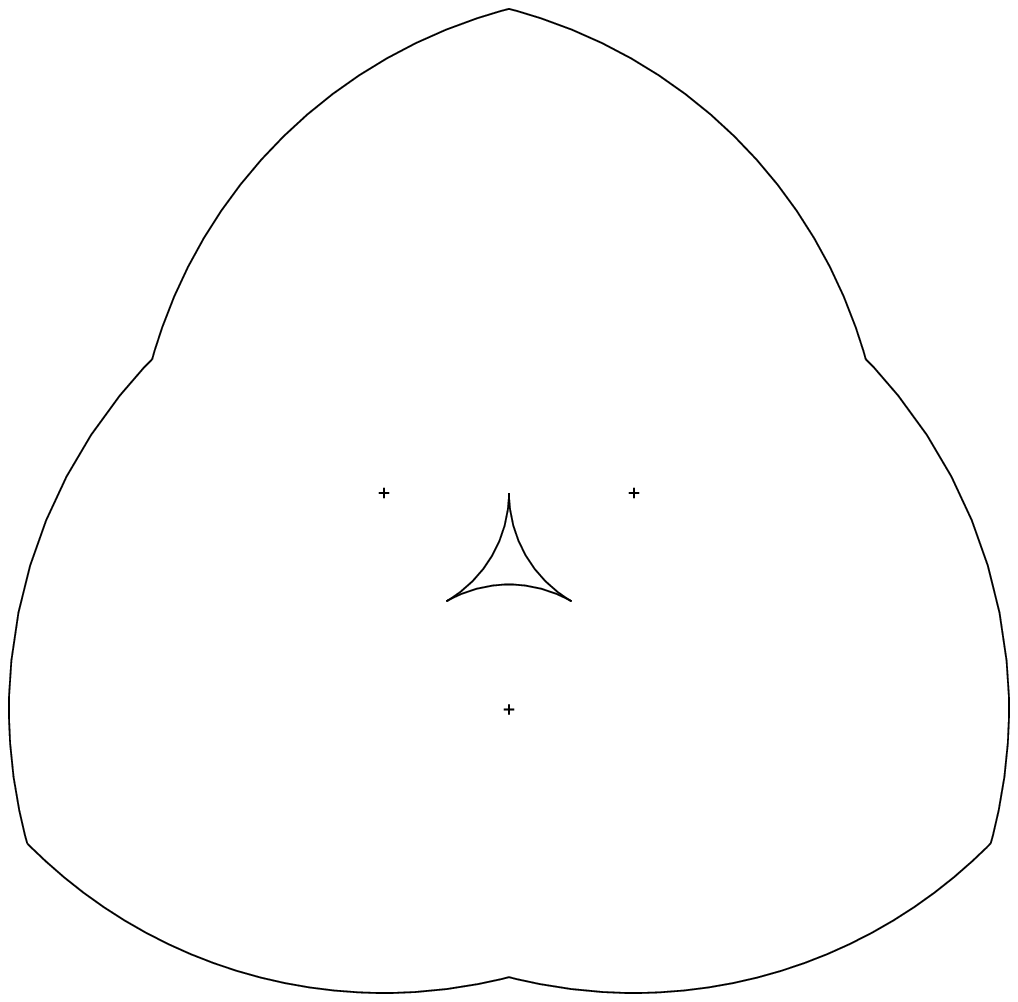,scale=0.4} \\
\parbox[t]{5.5cm}
{Figure 3. Another example for $\rho=1/2$.}
\hfill
\parbox[t]{5.5cm}
{Figure 4. A further example for $\rho=1/2$.}
\vspace{1mm}

A further possibility is an envelope of a water-line consisting of three arcs
which yields a boundary consisting of six arcs as shown in figure 4. In this
case $s(\phi)$ is piecewise constant with alternating signs.

\section{The case $\rho\not=\frac 12$. Linear Theory}

We continue our consideration on floating rods for densities $\rho\not=1/2$. We
describe the boundary in terms of polar coordinates $r(\psi)$. Each point of
the boundary can be at the water-level in two orientations given by $\phi$ and
$\chi$ (figure 5a), thus obeying
\bea
x(\psi) &=& r(\psi) \cos\psi
= x_M(\phi)+l\cos\phi = x_M(\chi)-l\cos\chi, \\
y(\psi) &=& r(\psi) \sin\psi
= y_M(\phi)+l\sin\phi = y_M(\chi)-l\sin\chi.
\eea

\epsfig{file=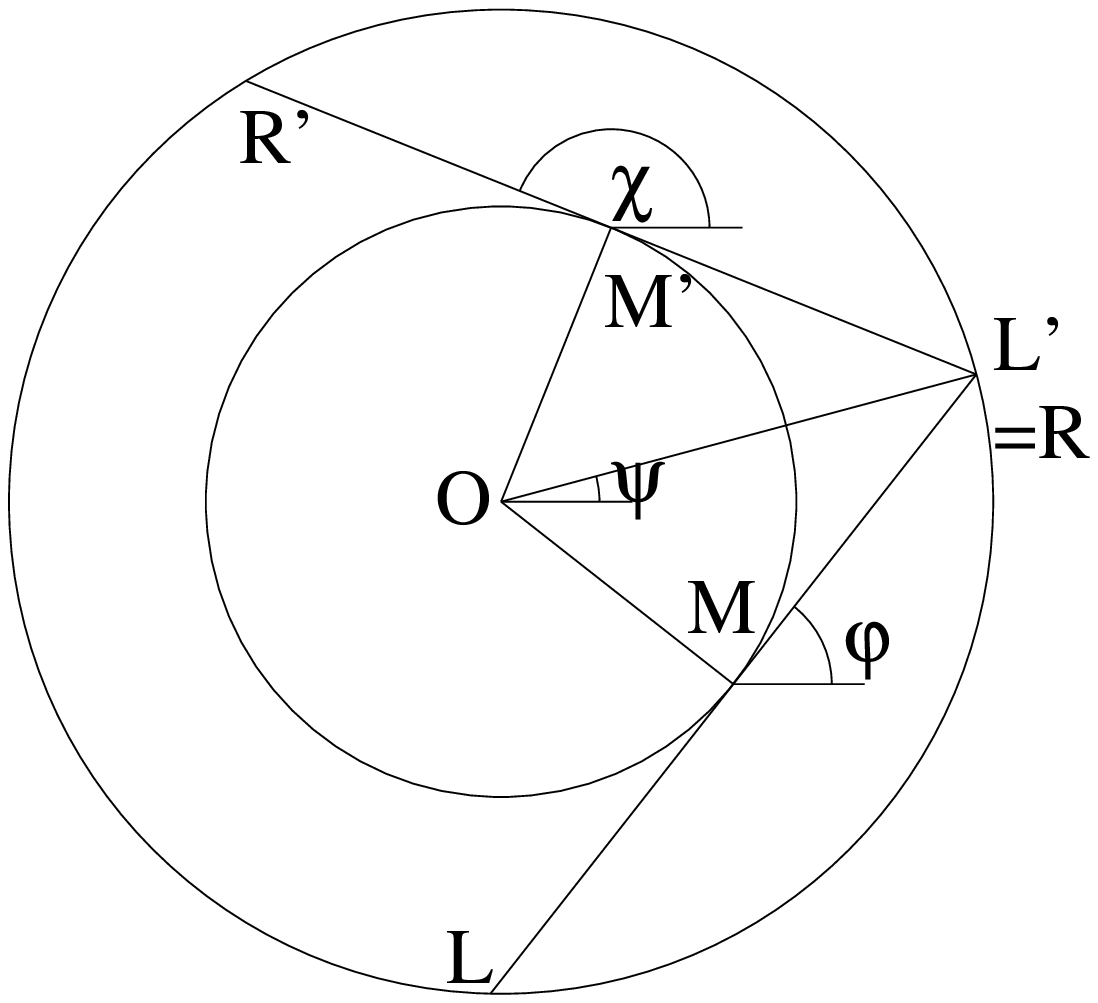,scale=0.5}
\hfill
\epsfig{file=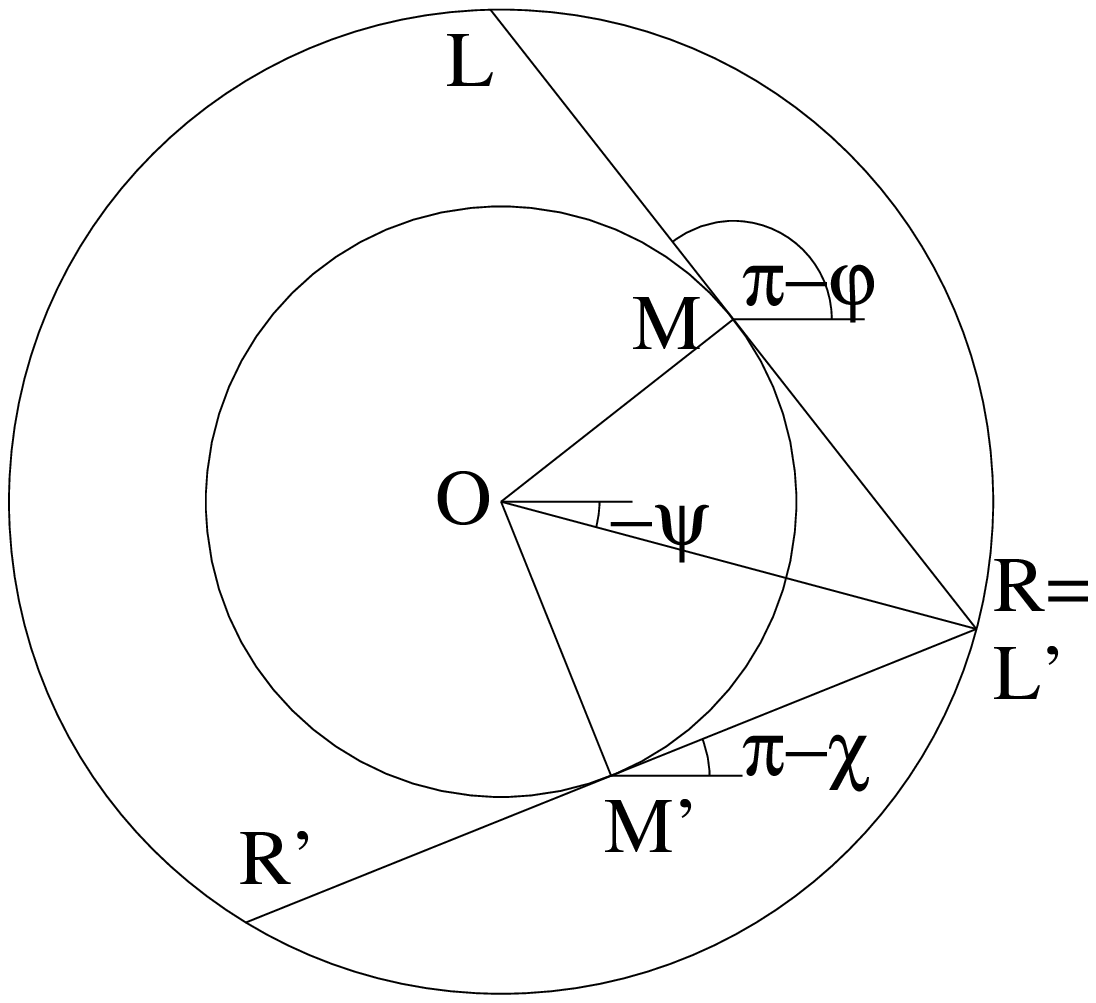,scale=0.5}

\parbox[t]{5.5cm}{Figure 5a. The two waterlines meeting at R=L'.}
\hfill
\parbox[t]{5.5cm}{Figure 5b. The picture reflected at the x-axis.}
\vspace{2mm}

We consider $\phi$ and $\chi$ to be functions of $\psi$ and differentiate
these equations with respect to $\psi$,
\bea
r' \cos\psi-r\sin\psi &=& \phi'(s(\phi)\cos\phi-l\sin\phi)
=\chi'(s(\chi)\cos\chi+l\sin\chi), \\
r' \sin\psi+r\cos\psi &=& \phi'(s(\phi)\sin\phi+l\cos\phi)
=\chi'(s(\chi)\sin\chi-l\cos\chi).
\eea
Multiplying these equations by $\cos\psi$ and $\sin\psi$ and adding or
subtracting them we obtain
\bea
r' &=& \phi'(s(\phi)\cos(\psi-\phi)+l\sin(\psi-\phi)) \nn
&=& \chi'(s(\chi)\cos(\chi-\psi)+l\sin(\chi-\psi)). \label{eq1a} \\
r&=& \phi'(s(\phi)\sin(\phi-\psi)+l\cos(\phi-\psi)) \nn
&=& \chi'(s(\chi)\sin(\chi-\psi)-l\cos(\chi-\psi)). \label{eq1b}
\eea

We first consider the trivial solution of a circular cross-section, where $r$
and $s$ are constant. For these we obtain
\bea
\phi=\psi+\delta_0, && \chi=\psi+\pi-\delta_0, \\
s=s_0=r_0\sin\delta_0, &&
\quad l=r_0\cos\delta_0, \quad r=r_0=\sqrt{s_0^2+l^2}.
\eea
For the circular case $\delta_0$ is the angle between the water-line $LR$ and
the radius $OR$ in figure 5a. 

In the next step we allow a variation of our four functions $r$, $s$, $\phi$
and $\chi$ in order to admit non-circular solutions. For this purpose we write
\be
r=r_0+\hr(\psi), \quad
s(\phi)=s_0+\hs(\phi), \quad
\phi=\psi+\delta_0+\hp(\psi), \quad
\chi=\psi+\pi-\delta_0+\hc(\psi).
\ee
Then the equations (\ref{eq1a}) to (\ref{eq1b}) read
\bea
\hr' &=& (1+\hp') (\hs(\psi+\delta_0+\hp)\cos(\delta_0+\hp)
-r_0\sin\hp), \label{eq2a} \\
&=& (1+\hc') (-\hs(\psi+\pi-\delta_0+\hc)\cos(\delta_0-\hc)
-r_0\sin\hc), \\
r_0+\hr &=& (1+\hp') (\hs(\psi+\delta_0+\hp)\sin(\delta_0+\hp)
+r_0\cos\hp), \\
&=& (1+\hc') (\hs(\psi+\pi-\delta_0+\hc)\sin(\delta_0-\hc)
+r_0\cos\hc). \label{eq2d}
\eea

\subsection{Linearization}

We assume, that the quantities $\hr$, $\hs$, $\hp$, $\hc$ are small quantities,
which can be expanded in powers of an expansion parameter $\epsilon$. Suppose
we have solved the equations (\ref{eq2a}) to (\ref{eq2d}) up to some order in
$\epsilon$ and wish to calculate the next order. Then by expanding these
equations in powers of $\epsilon$ we may extract the contributions in this new
order. The linear terms in this new order are unknown, the non-linear ones are
already known. We bring the linear terms on the left-hand side of the equations
(\ref{eq2a}) to (\ref{eq2d}) and the other terms on the right-hand side. This
yields the equations
\bea
\hr' +r_0\hp -\cos\delta_0 \hs(\psi+\delta_0) &=& I_1(\psi), \label{eq4a} \\
\hr' +r_0\hc +\cos\delta_0 \hs(\psi+\pi-\delta_0) &=& I_2(\psi), \label{eq4b}
\\ 
\hr -r_0\hp' - \sin\delta_0 \hs(\psi+\delta_0) &=& I_3(\psi), \label{eq4c} \\
\hr -r_0\hc' - \sin\delta_0 \hs(\psi+\pi-\delta_0) &=& I_4(\psi). \label{eq4d} 
\eea
where the non-linear terms are in the $I(\psi)$.
Let us expand now our unknowns in Fourier series,
\bea
\hr = \sum_k \hr_k \ex{\ie k\psi}, \quad
\hs(\phi) = \sum_k \hs_k \ex{\ie k\phi}, \quad
\hp = \sum_k \hp_k \ex{\ie k\psi}, \\
\hc = \sum_k \hc_k \ex{\ie k\psi}, \quad
I_n = \sum_k I_{n,k} \ex{\ie k\psi}.
\eea
Then our system of equations reads
\bea
\ie k \hr_k +r_0 \hp_k +C_{1,k} \hs_k = I_{1,k}, &&
C_{1,k}=-\cos\delta_0 \ex{\ie k\delta_0} \label{eq3a} \\
\ie k \hr_k +r_0 \hc_k +C_{2,k} \hs_k = I_{2,k}, &&
C_{2,k}=+\cos\delta_0 \ex{\ie k(\pi-\delta_0)} \\
\hr_k - \ie k r_0 \hp_k +C_{3,k} \hs_k = I_{3,k}, &&
C_{3,k}=-\sin\delta_0 \ex{\ie k\delta_0} \\
\hr_k - \ie k r_0 \hc_k +C_{4,k} \hs_k = I_{4,k}, &&
C_{4,k}=-\sin\delta_0 \ex{\ie k(\pi-\delta_0)}. \label{eq3d}
\eea
The determinant of this linear system of equations is $(1-k^2)\hC_k$ with
\bea
\hC_k &=& \ie k(C_{2,k}-C_{1,k}) + C_{4,k}-C_{3,k} \nn
&=& \left\{ \begin{array}{cc}
2\ie (k\cos\delta_0\cos(k\delta_0)+\sin\delta_0\sin(k\delta_0)) 
& k \mbox{ even} \\
-2(k\cos\delta_0 \sin(k\delta_0)-\sin\delta_0\cos(k\delta_0))
& k \mbox{ odd}.
\end{array} \right. \label{Chat}
\eea
Generally one has
\be
\hs_k = \frac{\hI_k}{\hC_k}, \quad
\hI_k = \ie k(I_{2,k}-I_{1,k}) + I_{4,k}-I_{3,k}.
\ee

If $\hC_k=0$, (which is always the case, if $k=0,\pm 1$) then a solution
exists only, if $\hI_k=0$. With the exception of $k=\pm 1$ the choice of
$\hs_k$ is arbitrary, and one obtains
\bea
(1-k^2)\hr_k &=& I_{3,k}+\ie kI_{1,k} - (\ie kC_{1,k}+C_{3,k})\hs_k \nn
&=& I_{4,k}+\ie kI_{3,k} - (\ie kC_{2,k}+C_{4,k})\hs_k \\
(1-k^2)r_0\hp_k &=& I_{1,k}-\ie kI_{3,k} - (C_{1,k}-\ie kC_{3,k})\hs_k \\
(1-k^2)r_0\hc_k &=& I_{2,k}-\ie kI_{4,k} - (C_{2,k}-\ie kC_{4,k})\hs_k
\eea
We consider now the cases $k=\pm 1$, $k=0$, and the special case, where $\hC_k$
vanishes for some other integer $k=p$.

\subsection{Translational Invariance and the case $k=\pm 1$}

In this subsection we show that the vanishing of the determinant for $k=\pm 1$
is related to the translational invariance of the problem. For $k=\pm 1$ we
have
\be
\hs_{\pm 1}=\int_0^{2\pi} \de\phi s(\phi) \ex{\pm\ie\phi} =
x_M(2\pi)-x_M(0) \pm\ie (y_M(2\pi)-y_M(0)) = 0.
\ee 
Thus we obtain only a solution, if
\be
I_{3,\pm 1} = \mp\ie I_{1,\pm 1}, \quad
I_{4,\pm 1} = \mp\ie I_{2,\pm 1}.
\ee
Then $\hp$ and $\hc$ can be expressed by
\be
r_0 \hp_{\pm 1} = \mp\ie \hr_{\pm 1} +I_{1,\pm 1}, \quad
r_0 \hc_{\pm 1} = \mp\ie \hr_{\pm 1} +I_{2,\pm 1}
\ee
with arbitrary $\hr$.

This arbitrary choice of $\hr_{\pm 1}$ is related to the translational
invariance of the problem. We are free to move the system by 
$(\delta x,\delta y)$. Then also the angle $\psi$ and the distance $r$ from
the origin will change,
\bea
\bar x(\psi+\delta\psi) = \bar r(\psi+\delta\psi) \cos(\psi+\delta\psi)
&=& x(\psi)+\delta x = r(\psi)\cos\psi + \delta x, \\
\bar y(\psi+\delta\psi) = \bar r(\psi+\delta\psi) \sin(\psi+\delta\psi)
&=& y(\psi) +\delta y = r(\psi)\sin\psi + \delta y.
\eea
With $\bar r(\psi)=r(\psi)+\delta r(\psi)$ one obtains
\bea
(r(\psi+\delta\psi))+\delta r)\cos(\psi+\delta\psi) 
&=& r(\psi)\cos(\psi)+\delta x, \\
(r(\psi+\delta\psi))+\delta r)\sin(\psi+\delta\psi) 
&=& r(\psi)\sin(\psi)+\delta y.
\eea
and thus
\bea
\delta x &=& (r'\delta\psi+\delta r)\cos\psi - r\sin\psi \delta\psi, \\
\delta y &=& (r'\delta\psi+\delta r)\sin\psi + r\cos\psi \delta\psi,
\eea
which yields
\bea
r\delta\psi &=& \delta y \cos\psi - \delta x \sin\psi, \\
\delta r &=& \delta x \cos\psi + \delta y \sin\psi 
- \frac{r'}r (\delta y \cos\psi - \delta x \sin\psi).
\eea
Since we presently consider the circular solution, $r'=0$ holds, and
\be
\delta r = \ex{\ie\psi} \hr_1 + \ex{-\ie\psi} \hr_{-1}
\ee
with $\hr_{\pm 1} = \frac{\delta x \pm \ie\delta y}2$. Since we have shifted
the system parallel, $s(\phi)$ and $s(\chi)$ did not change, $s_{\pm 1}=0$,
but $\bar{\phi}(\psi+\delta\psi) = \phi(\psi)$, which yields
\be
\delta\phi=-\phi'\delta\psi
=-\frac{\delta y}r \cos\psi + \frac{\delta x}r \sin\psi
=-\frac{\ie \hr_1}r \ex{\ie\psi} + \frac{\ie \hr_{-1}}r \ex{-\ie\psi}.
\ee
One obtains the same result for $\delta\chi$. This is in agreement with the
homogeneous equations (\ref{eq3a}) to (\ref{eq3d}) for $k=\pm 1$.

\subsection{Variation of the density and the case $k=0$}

A small but constant variation of $\delta_0 \rightarrow \delta_0 +\alpha$
yields for constant $l=r_0\cos\delta_0$
\be
\hr_0 \cos\delta_0 - r_0\alpha\sin\delta_0 =0.
\ee
This variation of $\delta_0$ corresponds to a variation of the density
$\rho$. For $s$, $\phi$ and $\chi$ we obtain
\be
\hs_0=\hr_0\sin\delta_0 + r_0\alpha\cos\delta_0, \quad
\hp_0=\alpha, \quad \hc_0=-\alpha.
\ee
Elimination of $\alpha$ yields
\be
\hr_0=\sin\delta_0\hs_0, \quad r_0\hp_0=\cos\delta_0\hs_0, \quad 
r_0\hc_0=-\cos\delta_0\hs_0
\ee
in agreement with the homogeneous solution of (\ref{eq3a}) to (\ref{eq3d}) for
$k=0$.

\subsection{The case $\hC_p=0$, $p\not=0,\pm 1$}

If $\hC_p(\delta_0)$ equals 0 for another $p$, then again we have a homogeneous
solution. This allows us to find non-circular solutions. Let us first
consider the zeroes of $\hC_k(\delta_0)$. We find, the $\hC$ vanishes for
$\cos\delta_0=0$. It turns out, that this is not a single zero, but a
three-fold zero. Since moreover $\hC_k(-\delta_0)=(-)^k \hC_k(\delta_0)$, one
finds, that $\hC_k$ is $\cos^3\delta_0$ times a polynomial of order $k-2$ in
$\cos\delta_0$ for even $k$, and $\sin\delta_0 \cos^3\delta_0$ times a
polynomial of order $k-3$ in $\cos\delta_0$ for odd $k$. Moreover one has
$C_k(\pi-\delta_0)=(-)^{k-1}C_k(\delta_0)$. Thus the polynomials are even in
$\cos\delta_0$ or simply polynomials in $\cos(2\delta_0)$. With the
abbreviations
\be
\cd=\cos\delta_0, \quad \sd=\sin\delta_0, \quad \cdz=\cos(2\delta_0)
\ee
we write
\bea
\hC_{2k+2} &=& 4\ie\cd^3 \, P_k(\cdz), \\
\hC_{2k+3} &=& -16\sd\cd^3 \, Q_k(\cdz)
\eea
and obtain
\bea
P_0 = 1, &&
Q_0 = 1, \\
P_1 = 6\cdz-4, &&
Q_1 = 4\cdz-1, \\
P_2 = 20\cdz^2-16\cdz+1, &&
Q_2 = 12\cdz^2-4\cdz-2, \\
P_3 = 56\cdz^3-48\cdz^2-12\cdz+8, &&
Q_3 = 32\cdz^3-12\cdz^2-12\cdz+2.
\eea
Limits for the zeroes of $\hC_p$ can be easily given, since the sign of
$\hC_p$ can be easily determined, if $\delta_0$ assumes integer multiples of
$\pi/(2p)$. Thus for even $p$ one has zeroes $\delta_{0,l}$ obeying
$\frac{\pi}p(l-\frac 12)<\delta_{0,l}<\frac{\pi}pl$ for $l=1,2,...\frac p2-1$.
For odd $p$ one has zeroes  
$\frac{\pi}pl<\delta_{0,l}<\frac{\pi}p(l+\frac 12)$ for 
$l=1,2,...\frac{p-3}2$.

The density corresponding to the angle $\delta_0$ is
\be
\rho_0 = \frac 12 - \frac{\delta_0}{\pi} - \frac{\sin(2\delta_0)}{2\pi}.
\ee
Thus in first order in $\hs$ we may distort the body from its circular shape by
adding a distortion described by
\be
\hs(\phi) = \hs_p \ex{\ie p\phi} + \hs_{-p} \ex{-\ie p\phi}.
\ee
and obtain a non-circular body which floats in any orientation. In
lowest order of the distortion it has density $\rho_0$. The angles and
densities for $p = 4 ... 9$ are listed below.

\begin{tabular}{rrrr}
 $p$ & $l$ & $\delta_0 [ ^0]$ & $\rho_0$ \\
  4 & 1 & 24.095 &  0.24751 \\
  5 & 1 & 37.761 &  0.13611 \\
  6 & 1 & 15.439 &  0.33255 \\
  6 & 2 & 46.670 &  0.08184 \\
  7 & 1 & 26.291 &  0.22753 \\
  7 & 2 & 52.959 &  0.05273 \\
  8 & 1 & 11.431 &  0.37466 \\
  8 & 2 & 34.361 &  0.16080 \\
  8 & 3 & 57.645 &  0.03585 \\
  9 & 1 & 20.261 &  0.28403 \\
  9 & 2 & 40.605 &  0.11713 \\
  9 & 3 & 61.273 &  0.02543
\end{tabular}
\vspace{1mm}

Only the solutions with positive $\delta_0$ are listed. Besides them the
angles $-\delta_0$ and the densities $1-\rho_0$ are also solutions. For odd
$p$ one has in addition the solution $\delta_0=0$ and $\rho_0=1/2$. In total
we have $p-2$ solutions $\delta_0$ for given $p$ without the unphysical
solutions $\cos\delta_0=0$ (see the discussion after eq. \ref{a}). Thus with
increasing $p$ the densities become arbitrarily dense.

We may also add some $\hs_0$, which however will not change the circular shape
but the density in first order in $\hs_0$.

\section{Higher Orders in the Distortion}

Our strategy will now be to calculate the contributions to $\hI_k$ in higher
orders in the distortion. As long as $k\not=p$, this will yield
contributions to $\hs_k$ which depend in higher orders on $\hs_0$ and
$\hs_{\pm p}$, but we have also to make sure, that $\hI_0$ and $\hI_{\pm p}$
vanish in all orders in the distortion.

\subsection{Rotational and Reflection Invariance}

The non-circular solution we have found in first order in the distortion has
a $p$-fold rotation axis and mirror symmetry. In higher orders in the
distortion we will obtain higher harmonics and quite generally we can expand
$\psi-\delta_0$, $\chi-\pi+\delta_0$, $s$ and $r$ in a Fourier-series which
contains only terms proportional to $\ex{\ie np\psi}$ with integer $n$. The
reason is, that we wish to keep the $p$-fold rotation axis. Then only
components $\ex{\ie m\psi}$ are allowed, which for $\psi \rightarrow
\psi+2\pi/p$ reproduce themselves, thus $\ex{\ie m 2\pi/p}=1$ is required,
which implies, that $m$ is an integer multiple of $p$. For our expansion we
use now $s_p=u$ and $s_{-p}=v$ as expansion parameters
\bea
\frac{\hs_k}{r_0} &=& \left\{ \begin{array}{lr}
u & k=p \\
v & k=-p \\
\sum_{n1,n2} \cs{n_1,n_2} u^{n_1} v^{n_2} \delta_{n,n_1-n_2} & k=np \\
0 & {\rm otherwise} \end{array} \right. \\
\frac{\hr_k}{r_0} &=& \left\{ \begin{array}{lr}
\sum_{n_1,n_2}\crr{n_1,n_2} u^{n_1} v^{n_2} \delta_{n,n_1-n_2} & k=np \\
0 & {\rm otherwise} \end{array} \right.
\eea
and for $\hp$ and $\hc$ with coefficients $\cp{}$ and $\cc{}$ similarly as for
$\hr$ with $\crr{}$. Thus we have
\bea
r(\psi) &=& r_0 \sum_{m,n} \crr{m,n} u^{m}v^{n} \ex{\ie p(m-n)\psi}, \\
s(\phi) &=& r_0 \sum_{m,n} \cs{m,n} u^{m}v^{n} \ex{\ie p(m-n)\phi},
\eea
similarly for $\phi(\psi)$ and $\chi(\psi)$.

Secondly we require for our solution the mirror symmetry, which holds in
first order in the distortion. If the x-axis is the axis for the mirror
symmetry of the cross-section, then (fig. 5b)
\be
r(\psi) = r(-\psi), \quad s(\phi) = s(\pi-\phi), \quad 
\phi(\psi)=\pi-\chi(-\psi).
\ee
In first order in $u$ and $v$ we have
\be
\hs(\phi) = u\ex{\ie p\phi} + v\ex{-\ie p\phi}
=\hs(\pi-\phi) = u \ex{\ie p(\pi-\phi)} + v\ex{-\ie p(\pi-\phi)}.
\ee
Thus $u$ and $v$ are connected by $v=(-)^p u$. Moreover we require $\hs$ to be
real, which implies $v=u^*$. Thus for even $p$ we have real $u$ and $v$,
whereas for odd $p$ they are imaginary. We now obtain
\bea
r(\psi) &=& r_0 \sum_{m,n} \crr{m,n} u^{m+n}(-)^{np} \ex{\ie p(m-n)\psi} \nn
= r(-\psi) &=& r_0 \sum_{m,n} \crr{m,n} u^{m+n}(-)^{np} \ex{-\ie (m-n)\psi} \nn
&=& r_0 \sum_{n,m} \crr{n,m} u^{m+n}(-)^{mp} \ex{\ie p(m-n)\psi}.
\eea
Comparison of the coefficients for the same powers of $u$ and of
$\ex{\ie p \psi}$ yields
\be
\crr{m,n} = (-)^{(m-n)p} \crr{n,m}.
\ee
Similarly one obtains
\be
\cs{m,n}=\cs{n,m}, \quad \cc{m,n} = (-)^{(m-n)p+1} \cp{n,m}.
\ee
We also see that under this transformation eq.(\ref{eq4a}) becomes the
negative of (\ref{eq4b}) and eq.(\ref{eq4c}) becomes (\ref{eq4d}). In both
cases $\psi$ translates into $-\psi$. Thus we have
\be
I_{1,m,n}=(-)^{(m-n)p+1} I_{2,n,m}, \quad I_{3,m,n}=(-)^{(m-n)p} I_{4,n,m}.
\ee
Moreover one has
\be
C_{1,(m-n)p} = - C_{2,(n-m)p}, \quad C_{3,(m-n)p} =  C_{4,(n-m)p}.
\ee

\subsection{The Expansion}

In determining the expansion coefficients $\crr{}$, $\cs{}$, $\cp{}$, and
$\cc{}$ we proceed as follows. First we expand the equations (\ref{eq4a}) to
(\ref{eq4d}) in powers of $u^mv^n$. Then we solve the coefficient equations
with increasing $n+m$. From the four equations with coefficients $u^mv^n$ we
use the first three ones in order to express $\crr{m,n}$, $\cp{m,n}$, and
$\cc{m,n}$ by the coefficients with either smaller $m$ or $n$, which
appear in the expressions $I_{1..4,m,n}$ and by the coefficient $\cs{m,n}$. If
$|m-n|>1$, then we may use the fourth equation in order to determine
$\cs{m,n}$. If $m=n$, then due to the reflection symmetry
$I_{3,n,n}=I_{4,n,n}$ and thus $\hI=0$. Thus the four equations are linearly
dependent and $\cs{n,n}$ is undetermined at this stage. When we come to
$(m,n)=(n+1,n)$, then in general $\hI_{n+1,n}$ will not vanish. However it
depends on $\cs{n,n}$, which is not yet determined. $\cs{n,n}$ enters in the
form \be
\hI_{n+1,n} = ... + 2\ie \frac{\cd^3 p(p^2-1)}W \cs{n,n}
\ee
for even $p$. For odd $p$ the factor $\ie$ is absent. Thus $\cs{n,n}$ is
uniquely determined from $\hI_{n+1,n}=0$.
$\cs{n+1,n}$ is already fixed to be zero with the exception
of $n=0$, where we have normalized it to $\cs{1,0}=1$. Due to the reflection
symmetry the equations for $(n,n+1)$ are simultaneously solved. A problem could
occur, if not only $\hC_{\pm p}=\hC_0=0$, but also for some $m$ one would have
$\hC_{mp}=0$. In our calculations which we performed up to $m+n=7$, we did
not observe any dangerous denominators in the coefficients.
The denominators of $\crr{}$, $\cp{}$, and $\cc{}$ contained only powers of
$p^2$, $p^2-1$, and $W$ (see below), and the denominators of $\cs{}$ only
powers of $p^2$, despite of the more complicated expressions $\hC_{np}$, which
one might have expected in the denominators.

The expansion was performed by use of Maple. In order to implement the
condition $\hC_p=0$, eq.(\ref{Chat}) I used the representation
\bea
\left. \begin{array}{c}
\cos(p\delta_0)=-\sd /W, \\
\sin(p\delta_0)=p\cd /W
\end{array} \right\} && \mbox{for even }p \\
\left. \begin{array}{c}
\cos(p\delta_0)=p\cd /W, \\
\sin(p\delta_0)=\sd /W
\end{array} \right\} && \mbox{for odd }p \\
\mbox{with } W=\pm \sqrt{p^2\cd^2+\sd^2}, && \label{exk}
\eea
where I used the explicit representation of $W$ only at a very late stage.

Here I give the results for $r$, $s$, $\phi$, and $\chi$ up to order $u^mv^n$
with $m+n\le 2$.
For even $p$ I obtain
\bea
\frac r{r_0} &=& 1+\frac W{p^2-1} (u\ex{\ie p\psi}+v\ex{-\ie p\psi})
+\frac{(2p^2-1)W^2}{2(p^2-1)^2p^2} (u^2\ex{2\ie p\psi}+v^2\ex{-2\ie p\psi})
\nn
&-& \frac{p^4(5+3\cdz) -4p^2\cdz +\cdz-1}{4(p^2-1)^2p^2} uv +... \\
\frac s{r_0} &=& \cd + u\ex{\ie p\psi} + v\ex{-\ie p\psi}
+\frac{(4p^2-1)\sd}{2p^2} (u^2\ex{2\ie p\psi}+v^2\ex{-2\ie p\psi}) \nn
&+&\frac{\sd}{2p^2} uv + ... \\
\phi &=& \psi+\delta_0
-\frac{\ie(p\cd+\ie\sd)(\cd-\ie p\sd)}{W(p^2-1)} u\ex{\ie p\psi} \nn
&+& \frac{\ie(p\cd-\ie\sd)(\cd+\ie p\sd)}{W(p^2-1)} v\ex{-\ie p\psi} \nn
&-& \frac{(p^2\cd-2\ie p\sd+\cd)\sd(p\cd+\ie\sd)^2}{2W^2(p^2-1)p^2}
u^2\ex{2\ie p\psi} \nn
&-& \frac{(p^2\cd+2\ie p\sd+\cd)\sd(p\cd-\ie\sd)^2}{2W^2(p^2-1)p^2}
v^2\ex{-2\ie p\psi} \nn
&+& \frac{\sd\cd(5p^2-1)}{2(p^2-1)p^2} uv + ... \\
\chi &=& \psi+\pi-\delta_0
+\frac{\ie(p\cd+\ie\sd)(\cd-\ie p\sd)}{W(p^2-1)} v\ex{-\ie p\psi} \nn
&-& \frac{\ie(p\cd-\ie\sd)(\cd+\ie p\sd)}{W(p^2-1)} u\ex{\ie p\psi} \nn
&+& \frac{(p^2\cd-2\ie p\sd+\cd)\sd(p\cd+\ie\sd)^2}{2W^2(p^2-1)p^2}
v^2\ex{-2\ie p\psi} \nn
&+& \frac{(p^2\cd+2\ie p\sd+\cd)\sd(p\cd-\ie\sd)^2}{2W^2(p^2-1)p^2}
u^2\ex{2\ie p\psi} \nn
&-& \frac{\sd\cd(5p^2-1)}{2(p^2-1)p^2} uv + ... 
\eea
For odd $p$ one has to replace $u$ by $-\ie u$ and $v$ by $\ie v$ in $r$,
$\phi$ and $\chi$, but one has to leave the factors $u$ and $v$ unchanged
in $s$.

In a next step I have fixed the radius averaged over all angles to $\bar r$ and
further denoted the relative amplitude of the first harmonic by $2\epsilon$,
that is $r(\psi)=\bar r(1+2\epsilon\cos(p\psi))+$higher harmonics. Then to my
surprise I obtained a solution depending only on $\epsilon$ and $p$, but not
explicitely on $\delta_0$ up to seventh order in $\epsilon$. Here I report the
result up to fifth order
\bea
\frac{r(\psi)}{\bar r} &=& 1 + 2\epsilon \cos(p\psi) \nn
&+& 2\left(\frac{2p^2-1}{2p^2}\epsilon^2
+\frac{(p^2-1)(30p^6-47p^4+12p^2-3)}{48p^6}\epsilon^4 \right) \cos(2p\psi) \nn
&+&
2\left(-\frac{(p^2-1)(3p^4-14p^2+3)}{16p^4}\epsilon^3 \right. \nn
&+&
\left.
\frac{3(p^2-1)^2(9p^8+160p^6-94p^4+24p^2-3)}{256p^8}\epsilon^5
\right) \cos(3p\psi) \nn
&-& 2\frac{30p^8-77p^6+83p^4-33p^2+3}{48p^6}\epsilon^4 \cos(4p\psi) \nn
&+& 2 \frac{(p^2-1)(75p^{10}-1011p^8+1414p^6-974p^4+255p^2-15)}
{768p^8}\epsilon^5 \cos(5p\psi) \nn
&+& O(\epsilon^6).
\eea
If this expansion continues in higher orders without dependence on $\delta_0$,
which seems quite likely, then we have for fixed $p$ and $\epsilon$ a
non-circular shape which can float in any direction for $p-2$ different
densities. This is an intriguing result.

The densities are functions of $\epsilon$,
\bea
\pi\rho &=& \pi\rho_0
-\frac{(p^2-1)^2\sd\cd^3}{p^2W^2}\epsilon^2
- \frac{(p^2-1)^2\sd\cd^3 P_{\rho}}
{32p^6W^4}\epsilon^4 + O(\epsilon^6), \quad\quad\quad\quad\quad\quad \\
P_{\rho} &=& 3p^8-26p^6-20p^4-22p^2+1+\cdz(3p^8-24p^6+6p^4+16p^2-1).
\eea

The ratio of the part of the circumference below the water-level $a$ to the
total circumference $a_{\rm tot}$ also depends on $\epsilon$ 
\bea
\frac{2\pi a}{a_{\rm tot}} &=& \pi - 2\delta_0
-\frac{\sd\cd(p^2-1)^2}{p^2W^2} \epsilon^2
+\frac{\sd\cd(p^2-1)^2 P_{\rm a}}{64p^6W^4}\epsilon^4 +O(\epsilon^6), \\
P_{\rm a}&=& 13p^8+50p^6-12p^4+14p^2-1+\cdz(13p^8-6p^4-8p^2+1). \label{a}
\eea

For $\rho=1/2$ all determinants $(k^2-1)\hC_k$ vanish for odd $k$. This is the
source for the large variety of cross-sections at $\rho=1/2$. We may, however,
use the above equations, which were initially solved for general $\cd$ and
$\sd$ and finally insert $\sd=0$, $\cd=1$, and again we obtain the same
boundary (to seventh order in $\epsilon$).

Finally we may reconsider the
''solution'' $\cd=0$, $\sd=\pm 1$. In this special case formally the boundary
and the envelope of water-lines coincide. However, the parametrization is
different, since $r$ is given as a function of the polar angle $\psi$, whereas
$x_M$, $y_M$ are given as function of the direction of the tangent on the
curve. Despite the fact that there is no solution for $\rho=0$ and $\rho=1$
except the circle (in this case the height of the center of gravity above or
below the water-line has to be constant, which according to Montejano
\cite{Montejano} is only fulfilled by a circle), our equations are fulfilled
for arbitrary closed boundaries, since in this limit $l=0$ and thus always
$\phi=\chi$. That $\cd=0$ appears as a multiple solution is probably a
consequence of the fact, that the fictitious water-line is parallel to the
boundary and thus the coincidence of $R$ and $L'$ is fulfilled in higher
orders. We note that formal substitution of $\cd=0$ and $\sd=\pm 1$ again
yields the same boundary.

There remains the question of the convergency of the $\epsilon$-expansion. I
do not have a definite answer to this. However, if one describes the boundary
by $x_M$, $y_M$ for $\cd=0$, then it is apparent, that there is a unique
solution only as long as the cross-section is convex, since otherwise $x_M$,
$y_M$ are no longer unique functions of $\phi$. Thus the radius of convergency
cannot be larger then the maximal $\epsilon$ for a convex boundary, which can
be estimated $\epsilon\approx 1/(2p^2)$. From the numerical calculations it
seems likely, that this is the radius of convergency for $\cd=0$. For larger
$\cd$, that is for $\rho$ closer to $1/2$ the radius of convergency seems to be
larger.

We show now several examples. For given $p$ and $\epsilon$ we have plotted
points of the envelope of the water-line  for different $\rho$ in intervals of
$6^0 = \pi/30$ and the corresponding points on the boundary for various cases.
For odd $p$ I include the result for $\rho=1/2$. As long as 
$\epsilon\le 1/(2p^2)$, I also include the points for $\cd=0$, which lie on
the boundary. If one would include the points calculated for $\cd=0$ for
$\epsilon=1/p^2$, then one would see that they scatter around and do not fit
on a smooth boundary, indicating the non-convergency of the
$\epsilon$-expansion in this case. All points have been calculated from the
expansion of $x_M$, $y_M$ to order $\epsilon^7$.

\begin{center}
\epsfig{file=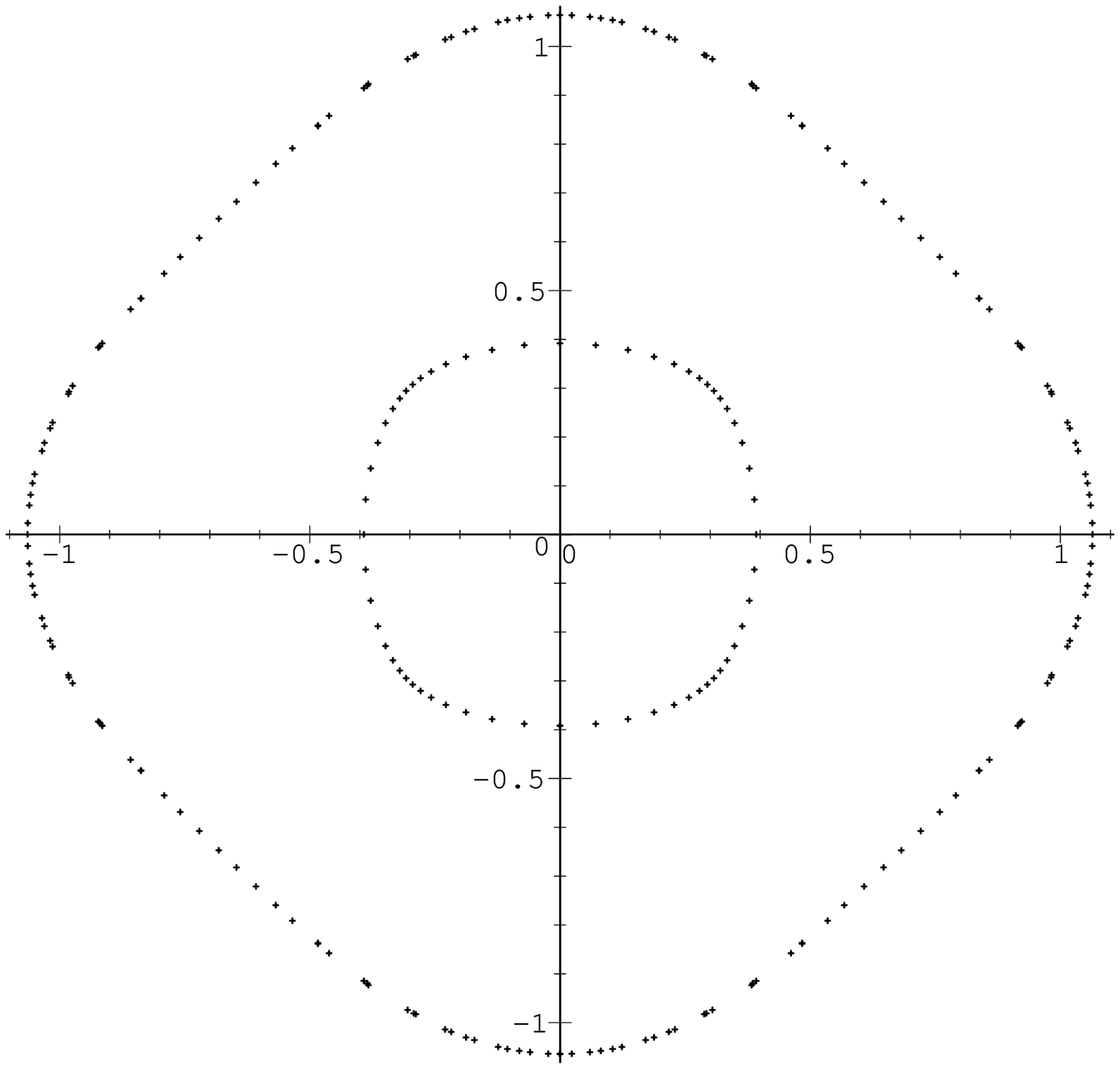,scale=0.4,angle=270} \\
Figure 6. $p=4$, $\epsilon=1/(2p^2)$. \\
\epsfig{file=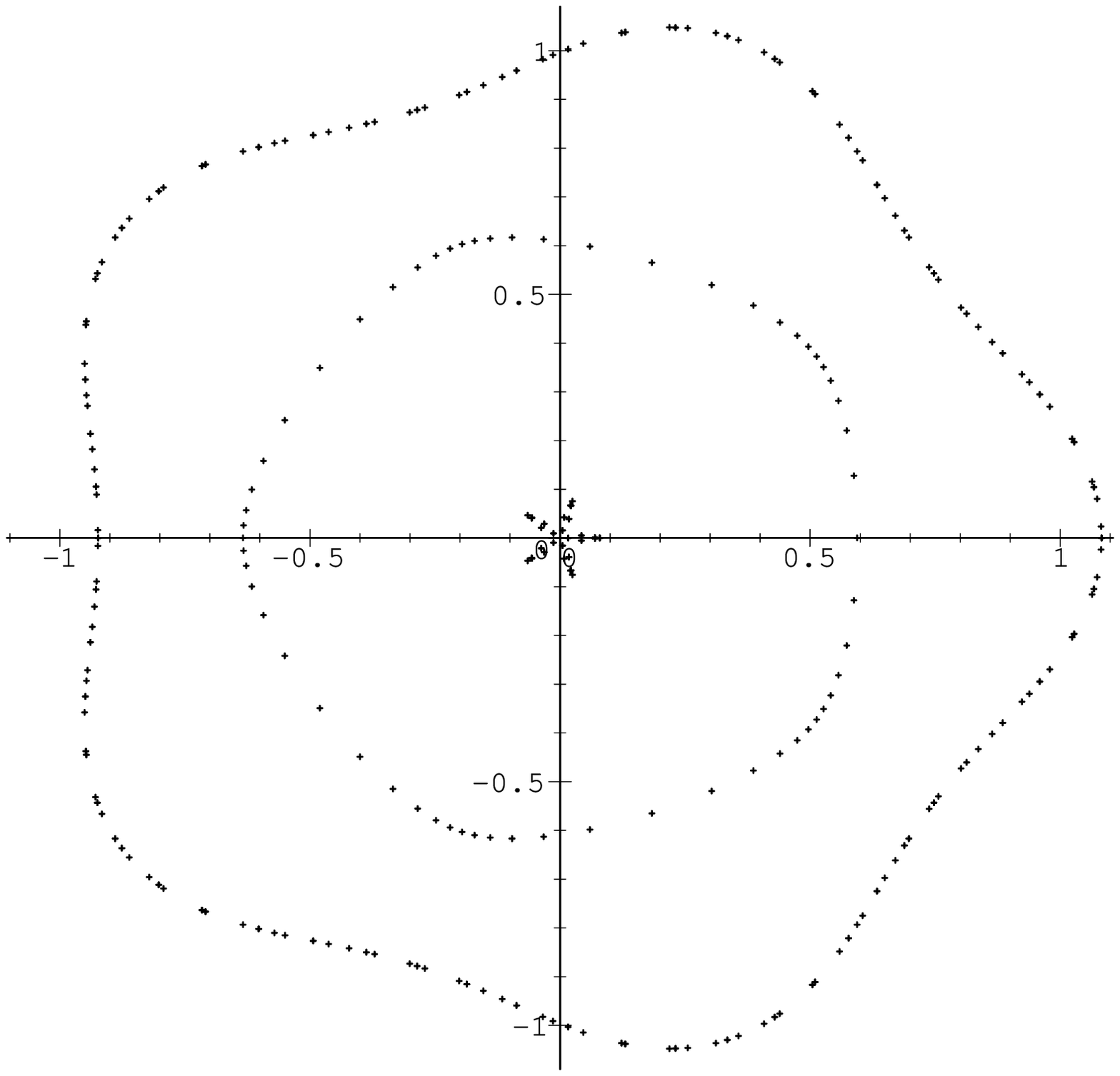,scale=0.4,angle=270} \\
Figure 7. $p=5$, $\epsilon=1/p^2$. \\
\epsfig{file=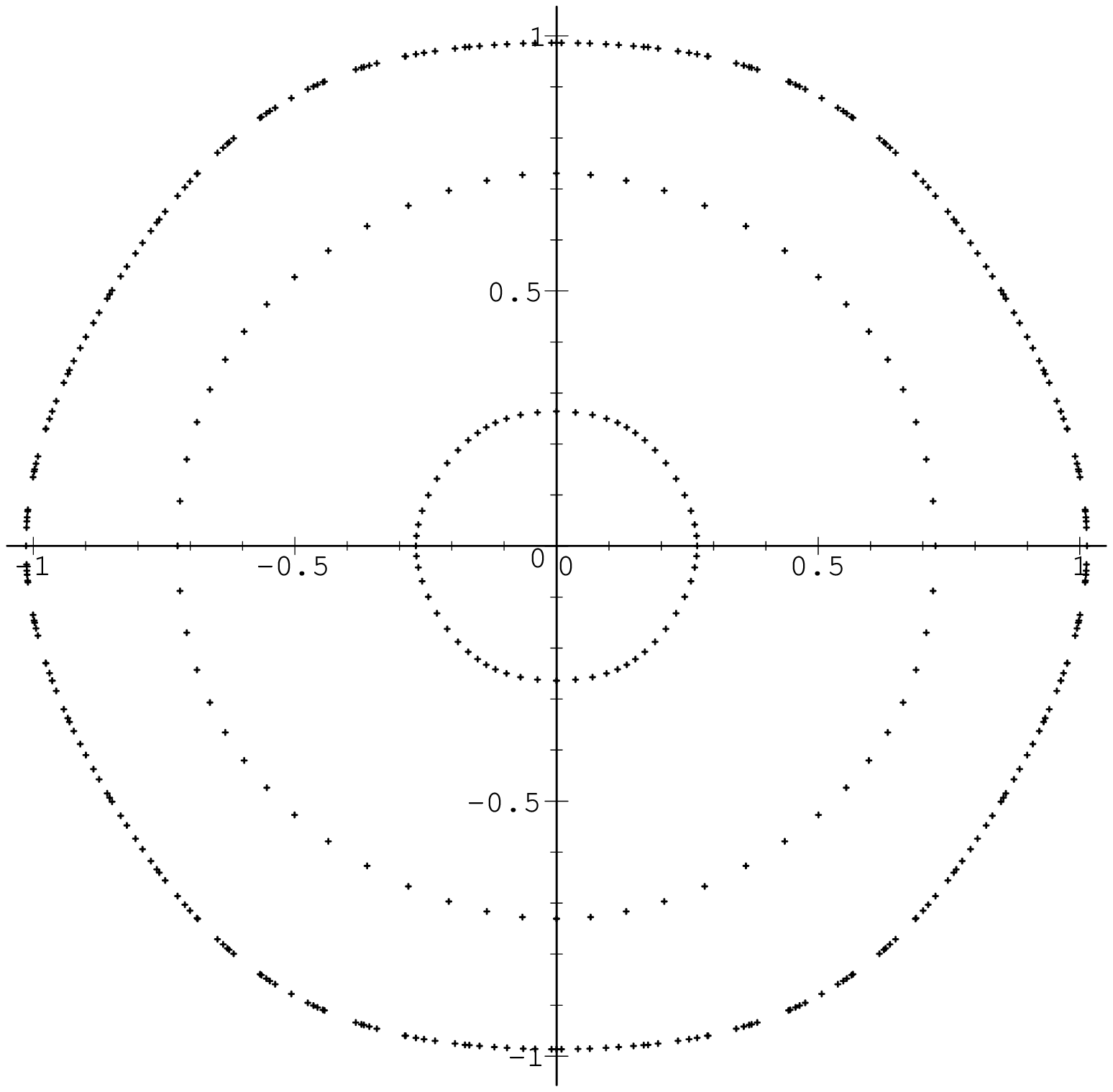,scale=0.4,angle=270} \\
Figure 8. $p=6$, $\epsilon=1/(4p^2)$. \\
\epsfig{file=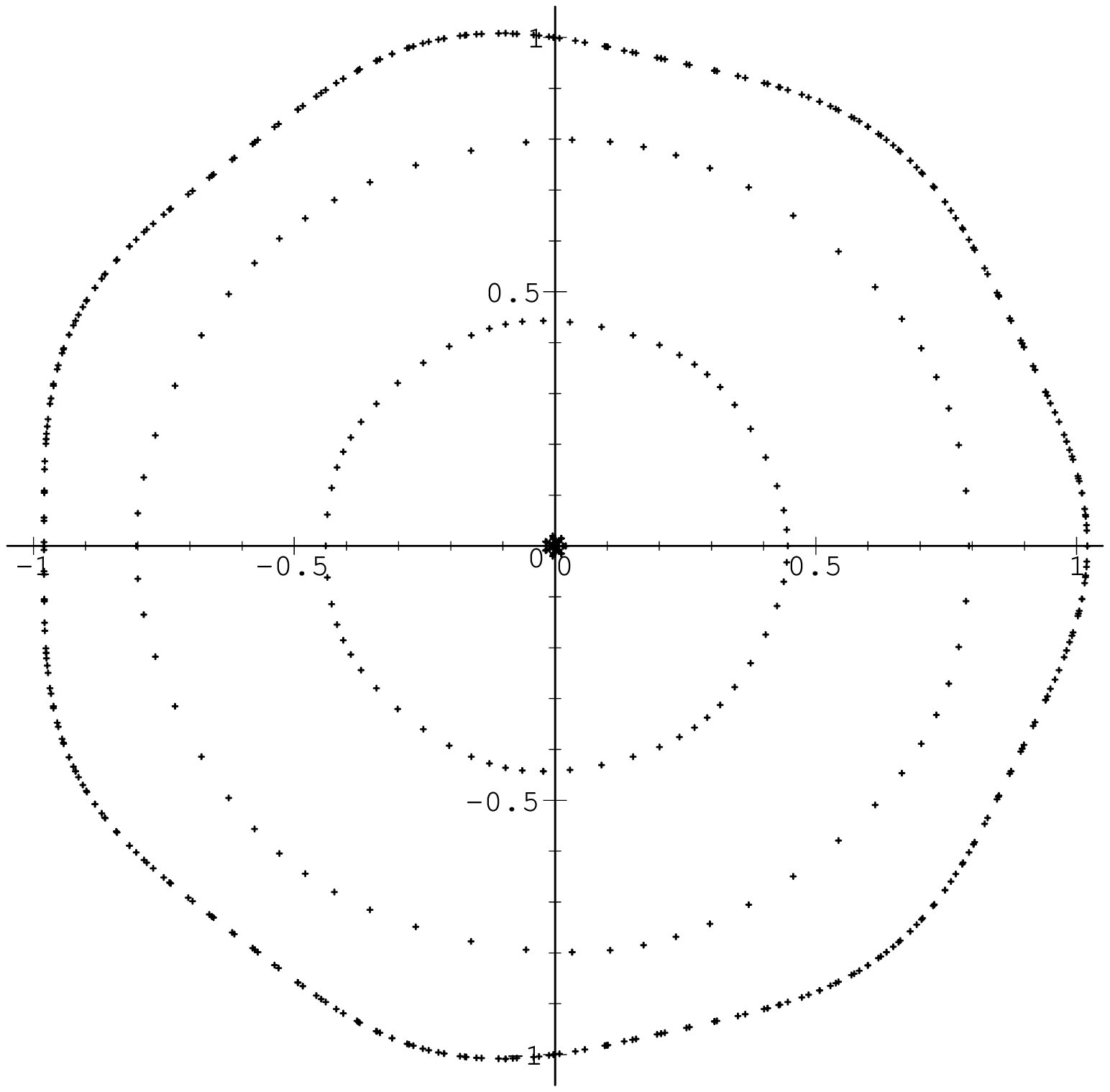,scale=0.4,angle=270} \\
Figure 9. $p=7$, $\epsilon=1/(2p^2)$. \\
\epsfig{file=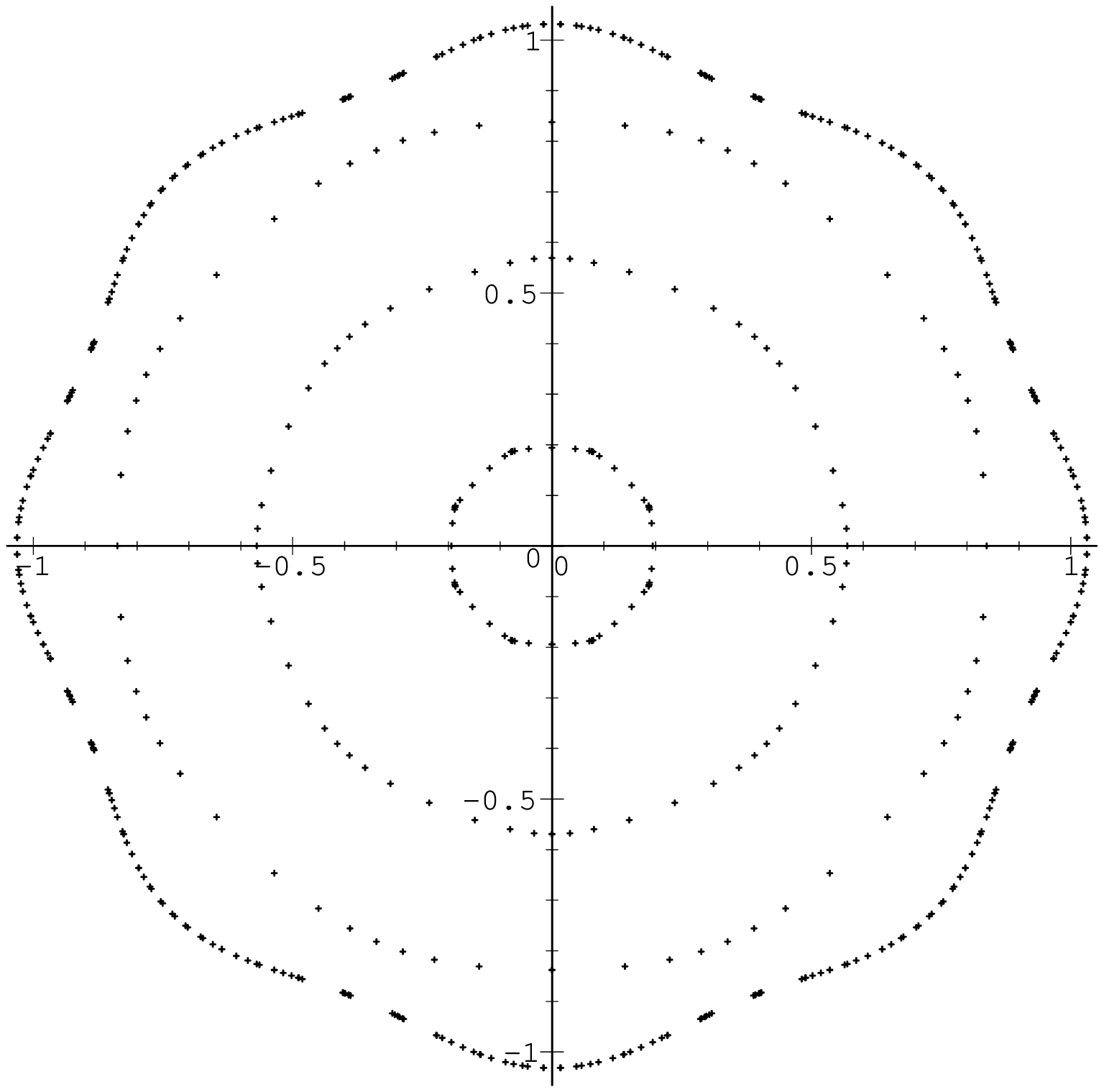,scale=0.4,angle=270} \\
Figure 10. $p=8$, $\epsilon=1/p^2$.
\end{center}

\section{Conclusion}

Here we have shown, that for densities different from $1/2$ it is possible to
have logs with non-circular cross-sections, which float indifferently in all
orientations around the axis. The cross-sections which are presented have a
$p$-fold rotation axis and $p$ mirror lines. For each
$p$ we have a one-parameter family of solutions at $p-2$ densities. {From} the
$\rho \rightarrow 1-\rho$ symmetry it is expected, that the solutions are
pairwise equal. However, the expansion in the deformation parameter $\epsilon$
yields $p-2$ equal solutions up to order $\epsilon^7$, from which I
conjecture, that one and the same manifold of cross-sections floats
indifferently for all $p-2$ densities. It would be very useful to obtain the
curves in a non-perturbative way, in order to verify or falsify the
conjecture.

We have obtained these solutions by deforming the circular solution. One may
also try to deform the solutions for $\rho=1/2$ or the solutions here obtained
by looking, if for some $\epsilon$ again a non-trivial deformation is possible
which might yield a solution of even lower symmetry.

{\bf Acknowledgment} I am indebted to Yacov Kantor for posing the problem in
his Physics Questions Series at http://star.tau.ac.il/QUIZ/ and for
correspondence.

\end{document}